# OCR quality affects perceived usefulness of historical newspaper clippings – a user study

Kimmo Kettunen[1], Heikki Keskustalo[2], Sanna Kumpulainen[2], Tuula Pääkkönen[3] and Juha Rautiainen[3]

[1] *University of Eastern Finland, School of Humanities, Finnish Language and Cultural Research, P.O. Box 111, 80101 Joensuu, Finland*
[2] *Tampere University, Faculty of Information Technology and Communication Sciences, Kalevantie 4, 33014 Tampereen yliopisto, Finland*
[3] *University of Helsinki, The National Library of Finland, Saimaankatu 6, 50100 Mikkeli, Finland*

**Abstract**

Effects of Optical Character Recognition (OCR) quality on historical information retrieval have so far been studied in data-oriented scenarios regarding the effectiveness of retrieval results. Such studies have either focused on the effects of artificially degraded OCR quality (see, e.g., [1-2]) or utilized test collections containing texts based on authentic low quality OCR data (see, e.g., [3]). In this paper the effects of OCR quality are studied in a user-oriented information retrieval setting. Thirty-two users evaluated subjectively query results of six topics each (out of 30 topics) based on pre-formulated queries using a simulated work task setting. To the best of our knowledge our simulated work task experiment is the first one showing empirically that users' subjective relevance assessments of retrieved documents are affected by a change in the quality of optically read text.

Users of historical newspaper collections have so far commented effects of OCR'ed data quality mainly in impressionistic ways, and controlled user environments for studying effects of OCR quality on users' relevance assessments of the retrieval results have so far been missing. To remedy this The National Library of Finland (NLF) set up an experimental query environment for the contents of one Finnish historical newspaper, *Uusi Suometar* 1869-1918, to be able to compare users' evaluation of search results of two different OCR qualities for digitized newspaper articles. The query interface was able to present the same underlying document for the user based on two alternatives: either based on the lower OCR quality, or based on the higher OCR quality, and the choice was randomized. The users did not know about quality differences in the article texts they evaluated.

The main result of the study is that improved optical character recognition quality affects perceived usefulness of historical newspaper articles significantly. The mean average evaluation score for the improved OCR results was 7.94% higher than the mean average evaluation score of the old OCR results.

## 1. Introduction

Digitized historical newspaper collections have been produced and increasingly used during the last two decades in different parts of the world, and both their usage and demand will increase in the future [4]. Access to these collections is important for different user groups, such as lay persons, teachers, journalists, and professional historians. Contents of the historical newspaper collections are produced using Optical Character Recognition, which has produced results of varying quality in the past. Although effects of OCR quality to search results have been evaluated in different settings, these studies have been performed either with artificially degraded OCR quality [1-2] or in IR laboratory-based experiments with original low quality OCR data [3]. Digital humanists have also evaluated the usability of historical newspaper query environments and commented possible problems caused by low OCR quality [5-6]. Low OCR quality has been also found to affect several activities during interactions with historical newspaper contents [7].

However, the effect of the OCR quality on perceived relevance of query results has not been studied yet. Therefore, we focus on real users in an experimental setting, where different quality OCR texts can be used at the same time and users perform the evaluations in a simulated work task setting. We aim at studying how the quality of OCR affects users' relevance assessment. To study this, we set up an experimental query environment for the contents of one Finnish historical newspaper, Uusi Suometar 1869-1918, with ca. 86 000 pages and ca. 306 M words. The collection includes ca. 1.45 million auto segmented different length articles, which we call *clippings*. The article database consisted of two versions of the same data: one with old, lower quality OCR and one with new, improved OCR.

In the interactive information retrieval experiment we used simulated work tasks [8-9] to trigger more naturalistic information needs. This allows individual test persons to assess the usefulness of the newspaper clippings with respect to their own interpretation. This increases the validity of the assessment. Further, we used graded relevance assessments. However, to increase control and repeatability during experimentation, we used pre-formulated, static queries.

The experimental query environment balloted between two different quality text versions presented for users and the users did not know about quality differences in the texts they evaluated. Thus, we could compare the subjective evaluations of the results. Our research question is whether different quality of the optical character recognition - old versus improved new - affects the perceived usefulness of the newspaper clippings.

## 2. Related research

Effects of sub-quality optical character recognition to efficiency of information retrieval have been studied earlier in different settings, and the results include both simulations, where quality of the text content has been tampered artificially, and usage of original Optical Character Recognition text. Actual user studies in a controlled query-environment, however, have been so far missing. Early simulated research settings include e.g., Taghva et al. [10], and Kantor and Voorhees [2], later ones Savoy and Naji [11], and Bazzo et al. [1], just to mention a few. The general result of these studies is that worse Optical Character Recognition quality lowers query results clearly. Most clear the effect of worse Optical Character Recognition is with short queries of a few words, where the query engine has less evidence for matching.

Järvelin et al. [3] report results of information retrieval in a laboratory style collection of digitized historical Finnish newspapers. Their collection consisted of 180 468 documents and 84 512 pages of newspapers, for which they had developed 56 search topics with graded relevance assessments. Results of the study show, that low level optical character recognition quality of the collection lowered search results clearly, even if heavy fuzzy-matching methods were used in query expansions to improve the results.

Van Strien et al. [12] suggest caution in trusting retrieval results of optically read text. They show, that both rankings of articles and number of returned articles from the query engine are affected by text quality. Traub et al. [13] show that better data quality decreases so called retrievability bias, which tends to bring certain documents as search result more often than others [14]. Chiron et al. [15]

show with respect to the French Gallica collection, that low frequency query words that contain frequent optical character error patterns have a higher risk to result in poor query results.

If we broaden scope and look at research outside information science, digital humanists have also paid attention to the problems of bad optical character recognition in digital historical newspaper collections. Jarlbrink and Snickars [5], for example, show how one digital Swedish newspaper collection, Aftonbladet 1830–1862, '*contains extreme amounts of noise: millions of misinterpreted words generated by OCR, and millions of texts re-edited by the auto-segmentation tool*'. Their main contribution is discussion of low-quality Optical Character Recognition and its effects on using digitized newspapers as research data. Pfanzelter et al. [6] describe user experiences and needs of digital humanities researchers with three digitized newspaper collections: Austrian ANNO (https://anno.onb.ac.at/), Finnish Digi (digi.kansalliskirjasto.fi), and French Gallica (https://gallica.bnf.fr/) and Retronews (https://gallica.bnf.fr/edit/und/retronews-0). Although their main concern in the paper is related to general functionality demands for interfaces of digitized newspaper collections, they report also experiences related to searchability of the collections. One of their general findings is that '*in some cases, the OCR quality is still very low. After identifying some major issues in this regard, the digital humanist team's reliance on (and trust in) some search results was very low*'.

Also, slightly differing opinions have been stated by digital humanities researchers. Strange et al. [16], for example, state that '*The cleaning was thus desirable but not essential*' referring by cleaning to correction of OCR errors in the digitized texts they were studying. Their comment was related to the word level accuracy of the texts – they did not consider a near optimal word level accuracy necessary. In their opinion a ca. 80% accuracy level was enough.

In an interactive information retrieval (IIR) setting simulated work task situations are used to trigger corresponding information needs [8-9]. An IIR setting requires three main facets [8]: i) potential users as test persons ii) application of dynamic and individual information needs and iii) use of multidimensional and dynamic relevance judgements. The interactive approach has the following four main advantages [8, 17]: first, it entails usage of cover-stories, which trigger information needs provoked by simulated work tasks. Second, the setting allows individual test persons to assess the usefulness of the newspaper clippings with respect to their own interpretation. Third, use of graded and multidimensional relevance assessments instead of binary and topical ones facilitates both control and repeatability during experimentation based on static queries. And finally, the setting enables the use of a realistic search interface with actual data.

## 3. Data and the experimental setting
### 3.1. Our newspaper data

Our search collection consists of the whole history of Uusi Suometar 1869–1918, ca. 86 000 pages and 306.8 million words [18]. Uusi Suometar was at the time of its publication one of the most important Finnish language newspapers in Finland, where newspapers were published in two languages, Finnish and Swedish. The original (old) optical character recognition for Uusi Suometar was performed using a line of ABBYY FineReader® products. Improved optical character recognition for the whole history of Uusi Suometar was achieved with Tesseract v.3.0.4.01. Improvement to the earlier quality in recognition of words is 15.3% units as a mean over the whole period. On average 83.6% of the words of the newspaper were recognized with automatic morphological analyzers, and the recognition rate varied from ca. 78 to 88% over the 49 years. For the old Optical Character Recognition mean word recognition rate was 68.3% [18]. Even if the improvement in Optical Character Recognition quality is considerable, the improved quality can still be challenging for information retrieval engines, especially with short queries and articles, where the information retrieval engine has less evidence for matching the query words and collection data in the engine's index [3].

Newspaper data at the National Library of Finland was originally scanned and recognized page by page without article structure information besides basic layout of the pages. For this study we used articles that were extracted automatically from the pages of Uusi Suometar using a trained machine learning model with software PIVAJ [19-21]. In the automatic segmentation process the collection of Uusi Suometar was divided into 1 459 068 articles with PIVAJ. The training of the PIVAJ model was based on 168 pages of manually marked newspaper page data that had different number of columns

(varying from 3 to 9). Kettunen et al. [21–22] reported success per centages of 67.9, 76.1, and 92.2 for an evaluation data set of 56 pages in three different evaluation scenarios based on Clausner et al. [23] using layout evaluation software from PRImA (https://www.primaresearch.org/).

In the article extraction of the whole history of Uusi Suometar article separation is far from optimal, and articles are perhaps best called automatically extracted clippings with varying length. In the search evaluation task, these clippings are documents that users search and evaluate. It should be emphasized, that the article segmentation that was producible for the whole history of Uusi Suometar is experimental and its quality will bring one layer of difficulty to the evaluation of search results. As Jarlbrink and Snickars [5] formulate it, auto segmentation tools create random texts, and borders of text snippets are fuzzy. This feature was informed to the users in the instructions of the search task.

## 3.2. The search environment

Participants of the evaluation task performed their task using the query engine Elastic search (https://www.elastic.co/), version 7.3.2, which is the background engine of the library's presentation system. Queries were performed in AND mode, where every query term is sought for in the documents. Hits of the search engine shown for the users needed to be at least 500 characters long to avoid very short text passages which would be hard to evaluate. The index of the newspaper collection's database is lemmatized, i.e., it contains base forms of the words, which is crucial as Finnish is a highly inflected language [3]. The articles of the newspapers had been extracted from the pages and stored as clippings. One clipping represented an article, and the search index contained the title and the textual contents of the specific article area taken from the OCR of ALTO XML[4] of the whole page, either from the original OCR page or from the re-OCR'ed page. The size of the original old OCR quality index is 9.82 Gb and the size of the re-OCR'ed index 9.04 Gb. Both indexes contain 1 459 068 clippings.

The query engine searched always for the results of queries in the new optical character recognition version of the database and ranked the results according to these. However, retrieved texts presented for reading were balloted in the two different optically read qualities of the same articles. Users of the query system were not aware of differences in the optical character recognition quality when they used the query environment.

Six pre-formulated queries for search and evaluation were presented for each user in the query form, one at a time. The query form interface is shown in Figure 1.

**Figure 1:** Screenshot of the query interface

Figure 1 shows the query interface after a pre-formulated query has been performed and 35 results retrieved, out of which 10 top results are shown for the user for evaluation. Text on the blue background on the top describes the topic and shows the pre-formulated query beneath in pink. The light purple rectangle below shows the beginning of the first query result. Relevance grading buttons 0–3 are on the right side of the rectangle. On the left, underneath the text snippet of the result, is the button for opening the clipping in its whole. The button also shows the character length of the clipping, 1943 characters, bottom line. Matches of the query words are highlighted in the snippet view and in the actual clipping view, which the participants used for evaluating relevance of the clippings.

---
[4] https://www.loc.gov/standards/alto/

### 3.3. Storing of the user session results

The article search and evaluation hackathon users had to log into the presentation system, so that their sessions could be stored. The users had previously agreed that their information is stored, and the log collected as little information about the individual user as possible. The structure of the result database is shown in Figure 2's Excel sheet.

**Figure 2:** Screenshot of the query log

The columns in the query log indicate the following data beginning from the left: A) query words B) session information C) number of the topic D) optical character recognition quality in the results (0 for the old and 1 for the new) E) user id F) role of the user (student or teacher) G) id number of the result clipping H) user-given evaluation result on the scale of 0-3 I) date and time of the session J) size of the clipping in characters K) rank (1-10) of the result clipping in the result list.

The interactive information retrieval system balloted the topics for each user, and out of the 32 users' work we got 1861 evaluations. This means that some of the users did not finish all their tasks, as the total number should have been 1920 (6 topics * top-10 evaluations * 32 users).

The clippings the users evaluated were of varying length. We had set a minimum length of 500 characters for the results to be shown for users, but no maximum length. The mean length of the clippings in all the evaluated results was 5467 characters.

### 3.4. Participants and their instructions

To perform the study, we recruited 32 participants for the evaluation task. The student users for the evaluation task were recruited from the courses *Information Retrieval and Language Technology* and *Information Retrieval Methods* at the Tampere University, Faculty of Information Technology and Communication Sciences. Three teachers of information science also participated in the evaluation task. Choice of the participants was based mainly on the ease of getting a large enough group to perform the tasks. We did not have access to a large enough group of historians with suitable search skills. We collected the information whether the users were students or teachers of information science (cf. Figure 2), but did not collect data about any other user qualities.

The participants were given background information, that their simulated task was to use the information retrieval system of digitized newspaper clippings to write an article about historical events in Finland or around the world during 1869-1918. Participants were given a one-page instruction leaflet which described the information retrieval task. The leaflet described the general idea of the task and retrieval session, gave them the back-grounding simulated work task story, and explained the graded evaluation scale of 0–3. The evaluation instructions advice the participant to consider how well the clipping helps accomplishing the task described in the background story, thus going beyond pure topical relevance assessment. Participants were guided to perform six queries. The queries were pre-formulated to increase control over the research setting and to guarantee the comparison between the users. The queries and the topic selection process are described in Appendix 1. Translations of the background story and the description of the graded relevance scale used in the evaluation task are in Table 1.

**Table 1.** The background story and evaluation instructions given to the participants

---

**The background story**

Imagine that you are writing an article related to topics in history of Finland or world history at the end of 19$^{th}$ century or the beginning of 20$^{th}$ century. Evaluate quality of the clippings you get as search results. Evaluate the quality of each clipping from the viewpoint, how it helps you to proceed with your article writing.

**Evaluation of the search results (graded relevance scale of 0-3)**

3. The clipping deals with the topic very broadly and its information content corresponds well with the task. The clipping helps well in accomplishing your task.
2. The clipping deals partially with the task or touches it. The content of the clipping helps to some extent in accomplishing your task.
1. The clipping does not deal with the actual topic but helps to find better search terms and to limit the topic somehow. It helps indirectly in accomplishing your task.
0. The clipping is wholly off topic and does not even help to formulate new queries. This clipping brings no benefit in accomplishing your task.

---

## 4. Results

Our research question was whether different quality of the optical character recognition (old versus improved new) affects the perceived usefulness of the newspaper clippings. We answer this by averaging the users' evaluation scores for all the evaluation results. On average 3.2 relevance assessments were made per clipping in the old Optical Character Recognition case, and 3.0 assessments in the case of the new improved Optical Character Recognition. Altogether, 961 and 900 assessments (correspondingly) were made in the two OCR qualities. Instead of these 1861 assessments there should have been 1920 assessments (32 users * 6 topics * 10 assessments), but evidently some users did not fully follow the instructions.

Mean averages for the evaluation scores over the whole query set for pre-formulated queries for the old OCR was 1.26 and for the new OCR 1.36. This reveals that the query results benefited from the improved optical character recognition. The mean average evaluation score for the improved OCR query results is 7.94% higher than the mean average score of the old OCR query results.

The difference in the effect of Optical Character Recognition quality on the relevance judgements was statistically significant (p=0.002, Wilcoxon's signed rank test [24]), when the relevance of the individual underlying documents was judged based on two possible levels of Optical Character Recognition quality. The difference in the overall effectiveness of retrieval (measured with mean average of cumulated gain among top-10 documents in the case of 30 topics), however, was not statistically significant (p=0.10, Wilcoxon's signed rank test).

Figure 3 shows mean averages for evaluations of the individual queries in the sessions with different quality optical character recognition.

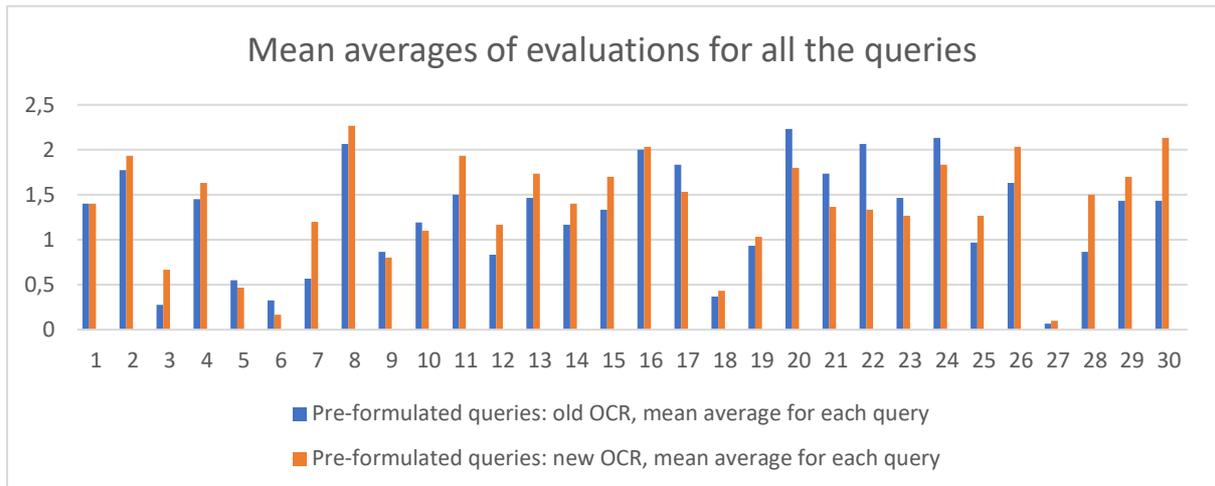

**Figure 3**: Mean averages of relevance scores for the top-10 clippings retrieved for all topics (N=30): graded relevance scale of 0-3 was used

Seven queries (#8, #16, #20, #22, #24, #26, and #30) got evaluations over relevance grade 2 with either OCR quality. Three queries, #3, #18, and #27, got low evaluations in both qualities. Query #6 got especially low evaluations in the new improved OCR.

Inspection of query-by-query results shows that improved OCR gains better mean evaluation scores in 19 cases out of 30. There is one tie (query #1) and 10 queries, where evaluations of old optical character recognition get better mean evaluation scores. This is depicted in detail in figure 4: the upwards pointing histograms show better mean relevance scores for improved OCR results, the downward pointing histograms show where improved OCR results have gained worse mean relevance scores.

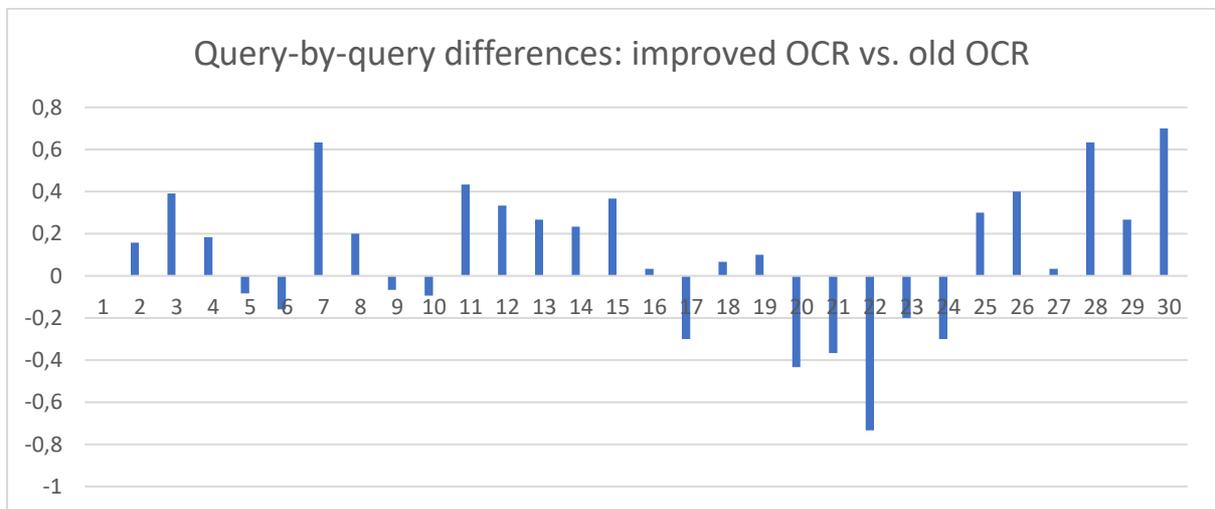

**Figure 4:** Query-by-query differences of relevance scores' mean averages for the top-10 clippings: graded relevance scale of 0-3 was used

Clearly over half of the query-by-query results with improved OCR were evaluated with higher mean scores than the old OCR results (19/30, 63.3%). One third of the queries got higher mean query-by-query evaluations with the old OCR. In general, the relevance level of the assessments is quite low, the mean being slightly over the lowest relevance level 1. Without closer inspection of the documents, it is not possible to deem, whether the low mean level of relevance assessments is due to the combination of OCR and clipping segmentation quality or due to other reasons.

A few notes with regards to our experiment are in order. First, our query environment implementation for the evaluation of two optically read text qualities is a first version of the system. As such it works well, but experience from user sessions showed that it has features that could be developed. We assumed that the user interface would take care of the number of queries and evaluations each user finished. However, some of the users acted against instructions and did not finish all the queries or evaluations in the sessions – the possibility of a user's premature quitting was not taken care of in the system.

Another possible development issue could be evaluation of the clippings' overall textual and segmentation quality by the users. Our article segmentation for the collection is experimental, and many of the clippings may be quite hard to read due to fuzzy boundaries: the clippings may contain text from adjacent segments, which affects evaluations. The users could also separately estimate the appropriateness of clipping boundaries, and presence of useful contents in the clipping.

## 5. Discussion and conclusion

To the best of our knowledge this is the first study showing empirically that the subjective relevance assessments of the test persons were affected by a change of quality of the optically read text presented to them. Earlier studies on the effects of Optical Character Recognition quality have been performed in data-oriented settings, using laboratory-style tests and artificially tampered data or they have described subjective experiences of users regarding the effects of Optical Character Recognition quality on their work.

The well-known simulated work task model used in interactive information retrieval [8] has been utilized in this study to answer the question of optical character recognition quality's effect on subjective relevance evaluation of retrieval results in a Finnish historical newspaper collection. We have shown that clear improvement in optical character recognition quality of documents leads to higher mean relevance evaluation scores in a simulated work task scenario. This means that perceived usefulness of historical newspaper clippings increases with better optical character recognition quality. The results imply that data-oriented scenarios of OCR quality effect evaluation should be augmented with more systematic user focused studies. By systemizing user-oriented studies for effects of OCR quality, new insights into the question can be achieved. Already the OCR quality has been found to have some effect on various information activities by causing extra work and some questions are raised amongst digital humanists about the reliability of research results based on the newspaper contents [7]. As we have shown, the simulated work task model offers a suitable paradigm for this kind of experiments.

Limitations of this study include our recruitment of test persons. Students and teachers of information research can be considered as experienced users of search engines, but on the other hand, they are not experts of history. Therefore, their evaluations of the resulting articles might differ from those of a group of historian users. A different group of users, be they historians or not, would evaluate results differently. However, considering the number of user tasks required in this experiment (30 topics), it was not feasible to recruit professional historians to act as test persons.

Although our results were achieved with one language, in one specific collection and with one user group, our method and model are generalizable to any language and can be evaluated with further users and different collections

Our results should be seen both in the context of the research methods necessary to apply in information retrieval and requirements of digital humanities scholars and lay users of the collections. These results bring more weight to both higher quality document need of digital humanists and efforts of improving quality of optical character recognition with new developments in software. Better quality of optically read historical documents should be strived for both for the sake of research and lay users.

## 6. Acknowledgements

This work was part of the NewsEye project, which has received funding from the European Union's Horizon 2020 research and innovation program under grant agreement No 770299. Faculty of Information Technology and Communication Sciences of the Tampere University took part in the arrangement of the query sessions and evaluation of the results as part of the Project EVOLUZ (#326616) financed by the Academy of Finland.
The query environment was implemented by Evident Ltd. (https://evident.fi/).

## 7. References

[1] G.T. Bazzo, G.A. Lorentz, V. D. Suarez, V.P. Moreira, Assessing the Impact of OCR Errors in Information Retrieval, in: J. Jose et al. (Eds.), Advances in Information Retrieval, ECIR 2020, Lecture Notes in Computer Science, vol 12036, Springer, Cham. doi: 10.1007/978-3-030-45442-5_13.

[2] P. B. Kantor, E.M. Voorhees, The TREC-5 confusion track: comparing retrieval methods for scanned text, Inf. Retrieval 2 (2000) 165–176. doi: 10.1023/A:1009902609570

[3] A. Järvelin, H. Keskustalo, E. Sormunen, M. Saastamoinen, K. Kettunen, Information retrieval from historical newspaper collections in highly inflectional languages: A query expansion approach, Journal of the Association for Information Science and Technology 67 (2016) 2928-2946. doi: 10.1002/asi.23379.

[4] M. H. Beals, Emily Bell, with contributions by Ryan Cordell, Paul Fyfe, Isabel Galina Russell, Tessa Hauswedell, Clemens Neudecker, Julianne Nyhan, Sebastian Padó, Miriam Peña Pimentel, Mila Oiva, Lara Rose, Hannu Salmi, Melissa Terras, and Lorella Viola, The Atlas of Digitised Newspapers and Metadata: Reports from Oceanic Exchanges, Loughborough, 2020. doi: 10.6084/m9.figshare.11560059.

[5] J. Jarlbrink, P. Snickars, Cultural heritage as digital noise: nineteenth century newspapers in the digital archive, Journal of Documentation 73 (2017) 1228-1243. doi: 10.1108/JD-09-2016-0106.

[6] E. Pfanzelter, S. Oberbichler, J. Marjanen, P.-C Langlais, S. Hechl Digital interfaces of historical newspapers: opportunities, restrictions and recommendations, The Journal of Data Mining & Digital Humanities (2021). doi: 10.46298/jdmdh.6121.

[7] E. Late, S. Kumpulainen, Interacting with digitised historical newspapers: understanding the use of digital surrogates as primary sources, Journal of Documentation (ahead-of-print). doi: 10.1108/JD-04-2021-0078.

[8] P. Borlund, Experimental Components for the Evaluation of Interactive Information Retrieval Systems, Journal of Documentation, 56 (2000) 71–90. doi: 10.1108/EUM0000000007110.

[9] P. Ingwersen, K. Järvelin, The Turn. Integration of Information Seeking and Retrieval in Context. Springer, 2005.

[10] K. Taghva, J. Borsack, J., A. Condit, Evaluation of Model-Based Retrieval Effectiveness with OCR Text, ACM Transactions on Information Systems 14 (1996) 64–93. doi: 10.1145/214174.214180.

[11] J. Savoy, N. Naji, Comparative Information Retrieval Evaluation for Scanned Documents, in: Proceedings of the 15th WSEAS International Conference on Computers, 2011, pp. 527–534. doi: 10.5555/2028299.2028394.

[12] D. van Strien, K. Beelen, M. C. Ardanuy, K. Hosseini, B. McGillivray, G. Colavizza, Assessing the Impact of OCR Quality on Downstream NLP Tasks, in: Proceedings of the 12th International Conference on Agents and Artificial Intelligence - Volume 1: ARTIDIGH, 2020, pp. 484-496. doi: 10.5220/0009169004840496.

[13] M.C Traub., J. van Ossenbruggen, L. Hardman, Impact Analysis of OCR Quality on Research Tasks in Digital Archives, in: S. Kapidakis, C. Mazurek C., M. Werla (Eds.), Research and Advanced Technology for Digital Libraries. TPDL 2015. Lecture Notes in Computer Science, vol 9316. Springer, Cham, 2015. doi: 10.1007/978-3-319-24592-8_19.

[14] L. Azzopardi, V. Vinay, Retrievability: an evaluation measure for higher order information access tasks, in: Proceedings of the 17th ACM conference on Information and knowledge management (CIKM '08), 2008. doi: 10.1145/1458082.1458157.
[15] G. Chiron, A. Doucet, M. Coustaty, M. Visani, J. Moreux, Impact of OCR Errors on the Use of Digital Libraries: Towards a Better Access to Information. ACM/IEEE Joint Conference on Digital Libraries (JCDL), Toronto, ON, 2017, pp. 1-4. doi: 10.1109/JCDL.2017.7991582.
[16] C. Strange, D. McNamara, J. Wodak, I. Wood, Mining for the Meanings of a Murder: The Impact of OCR Quality on the Use of Digitized Historical Newspapers, Digital Humanitites Quarterly 8 (2014). URL: http://www.digitalhumanities.org/dhq/vol/8/1/000168/000168.html.
[17] P. Borlund, P. Ingwersen, Measures of relative relevance and ranked half-life: performance indicators for interactive IR. in: SIGIR '98: Proceedings of the 21st annual international ACM SIGIR conference on Research and development in information retrieval, August 1998, pp. 324–331. doi: 10.1145/290941.291019.
[18] M. Koistinen, K. Kettunen, J. Kervinen, How to Improve Optical Character Recognition of Historical Finnish Newspapers Using Open-Source Tesseract OCR Engine – Final Notes on Development and Evaluation, in: Z. Vetulani, P. Paroubek, M. Kubis (Eds.), Human Language Technology. Challenges for Computer Science and Linguistics. LTC 2017. Lecture Notes in Computer Science, vol 12598, Springer, Cham, 2020. doi: 10.1007/978-3-030-66527-2_2.
[19] D. Hebert, T. Palfray, T. Nicolas, P. Tranouez, T. Paquet, PIVAJ: displaying and augmenting digitized newspapers on the web experimental feedback from the "Journal de Rouen" collection, in: Proceeding DATeCH 2014 Proceedings of the First International Conference on Digital Access to Textual Cultural Heritage, 2014, pp. 173–178. doi: 10.1145/2595188.2595217.
[20] D. Hebert, T. Palfray, T. Nicolas, P. Tranouez, T. Paquet, Automatic article extraction in old newspapers digitized collections, in: Proceeding DATeCH 2014 Proceedings of the First International Conference on Digital Access to Textual Cultural Heritage 2014, pp. 3–8. doi: 10.1145/2595188.2595195.
[21] K. Kettunen, T. Ruokolainen, E. Liukkonen, P. Tranouez, D. Anthelme, T. Paquet, Detecting Articles in a Digitized Finnish Historical Newspaper Collection 1771–1929: Early Results Using the PIVAJ Software, in: DATeCH2019: Proceedings of the 3rd International Conference on Digital Access to Textual Cultural Heritage May 2019, pp. 59–64. doi: 10.1145/3322905.3322911
[22] K. Kettunen, T. Pääkkönen, E. Liukkonen, Clipping the Page – Automatic Article Detection and Marking Software in Production of Newspaper Clippings of a Digitized Historical Journalistic Collection, in: A. Doucet, A. Isaac, K. Golub, T. Aalberg, A. Jatowt (Eds.), Digital Libraries for Open Knowledge 23rd International Conference on Theory and Practice of Digital Libraries, TPDL 2019, Oslo, Norway, September 9-12, 2019, Proceedings. Lecture Notes in Computer Science , no. 11799 , Springer Nature Switzerland , Basel, pp. 356-60 , TPDL 2019 , 09/09/2019. doi: 10.1007/978-3-030-30760-8_33.
[23] C. Clausner, S. Pletshacher, A. Antonacopoulos, Scenario Driven In-depth Performance Evaluation of Document Layout Analysis Methods, 2011 International Conference on Document Analysis and Recognition, 2011, pp. 1404-1408. doi: 10.1109/ICDAR.2011.282.
[24] W. B. Croft, D. Metzler, T. Strohman, Search Engines. Information Retrieval in Practice. Pearson, 2010.
[25] S. Zetterberg (Ed.), Suomen historian pikkujättiläinen, WSOY, 1989.
[26] S. Zetterberg (Ed.), Maailmanhistorian pikkujättiläinen, WSOY, 1988.

# Appendix 1. Topic creation and list of the pre-formulated queries

The topics were created using history timelines from two popular history encyclopedias: *Suomen historian pikkujättiläinen* [25] ('A small encyclopedia of Finnish history') and *Maailmanhistorian pikkujättiläinen* [26], ('A small encyclopedia of world history'). After finding suitable topics from the timelines of the encyclopedias, searches to the newspaper data base at digi.kansalliskirjasto.fi were performed to confirm that the database had enough hits related to the topic. During final creation of the query environment many original topics were abandoned, and new ones were created due to too few hits in the final article extraction database. Final topic descriptions were based on Finnish Wikipedia articles related to the topics. The topics cover the time frame of the historical collection of Uusi Suometar, beginning from 1870s and ending in 1918. First mentioned year in the topic descriptions is 1871, last 1918. Topics cover both domestic and foreign news, the share of domestic news being 21, and foreign 9. Demarcation line between foreign and domestic news is not always sharp, some topics could be classified as both. The mean length of the pre-formulated queries is 2.87 words.

| ID | Query in Finnish | Rough translation |
|---|---|---|
| 1 | Bobrikoffin murha 1904 | Murder of (Nikolai) Bobrikoff in 1904 |
| 2 | Postimanifesti 1890 | Postal manifest in 1890 |
| 3 | Nuorsuomalaisen puolueen perustaminen vuonna 1894 | Founding of the young Finns' party in 1894 |
| 4 | Helmikuun manifesti 1899 | The February manifest in 1899 |
| 5 | Eduskuntavaalit 1907 | Parliamentary elections in 1907 |
| 6 | Hannes Kolehmainen Tukholman olympialaisissa 1912 | Hannes Kolehmainen at the Stockholm Olympics in 1912 |
| 7 | Maailmansodan rauha 1918 | Peace of the WWI in 1918 |
| 8 | Nansenin matka pohjoisnavalle | Nansen's expedition to the North Pole |
| 9 | Lokakuun vallankumous Venäjällä 1917 | October revolution in Russia year 1917 |
| 10 | Saksan keisarikunta 1871 | The German Empire 1871 |
| 11 | Norjan itsenäisyys 1905 | Independence of Norway in 1905 |
| 12 | Tampereen valloitus 1918 | Conquest of Tampere in 1918 |
| 13 | Suomen kuningas Friedrich Karl | Karl Friedrich, the King of Finland |
| 14 | Tokoin senaatti 1917 | The senate of (Oskari) Tokoi |
| 15 | Tukholman olympialaiset 1912 | The Olympic games of Stockholm in 1912 |
| 16 | Maamieskoulu | Agricultural school |
| 17 | Laukon torpparilakko | Sharecroppers' strike in Laukko |
| 18 | Helsingin valtaus 1918 | Occupation of Helsinki in 1918 |
| 19 | Suomen itsenäisyys 1917 | Independence of Finland in 1917 |
| 20 | Espanjantauti | The Spanish flu |
| 21 | Viaporin kapina 1906 | Rebellion in Viapori in 1906 |
| 22 | Laulaja Aino Ackte | Singer Aino Ackte |
| 23 | Suomen laulu kuoro | The choir of Finnish song |
| 24 | Suomen Naisyhdistys | The Finnish Womens' association |
| 25 | Lontoon olympialaiset 1908 | London Olympics 1908 |
| 26 | Raitiotie Helsingissä | Tramway in Helsinki |
| 27 | J. L. Runebergin kuolema 1877 | Death of J.L. Runeberg in 1877 |
| 28 | Mannerheim valtionhoitajana 1918 | (General) Mannerheim as a regent in 1918 |
| 29 | Torpparilaki | The sharecropper law |
| 30 | Elinkeinovapaus 1879 | Freedom of livelihood in 1879 |